\documentclass[aps,pra,showpacs,twocolumn]{revtex4}
\usepackage{epsfig}
\begin{document}
\title{Continuous-variable quantum 
non-demolishing interaction at a distance}
\author{Radim Filip}
%\footnote{email:filip@optnw.upol.cz, tel:+420-68-5631572,
%fax:+420-68-5224246}
\affiliation{Department of Optics, Palack\' y University,\\
17. listopadu 50,  772~07 Olomouc, \\ Czech Republic}
\date{\today}
\begin{abstract}
A feasible setup of continuous-variable (CV) quantum non-demolishing (QND) interaction at a distance is proposed. 
If two distant experimentalists are able to locally perform identical QND interactions then the proposed 
realization requires only a single quantum channel and classical communication between them. 
A possible implementation of the proposed setup in recent quantum optical laboratories is discussed 
and an influence of Gaussian noise in the quantum channel on a quality of the implementation is analyzed. 
An efficient realization of the QND interaction at a distance can be a basic step to possible distributed quantum 
CV experiments between the distant laboratories.
\end{abstract}
\pacs{03.67.-a}
\maketitle

Quantum communication can be used to extend a local quantum interaction at a distance. It is a way 
how distant laboratories can directly collaborate on an experiment without moving of an apparatus from one laboratory to another
and perform distributively a joint quantum experiment. The distributed quantum experiment consists of local 
sub-experiments which can be linked together by quantum channels.   
An efficient experimental implementation of a basic interaction at a distance is an elementary step to build the distributed experiment.
Two distant experimentalists Alice and Bob can straightforwardly 
accomplish an arbitrary joint unitary operation ${\bf U}$ between two systems $A,B$ at a distance if they share a pair of quantum channels.
Alice transmits her system $A$ to Bob through the first quantum channel, 
Bob performs the operation ${\bf U}$ on the systems $A,B$ and subsequently transmits the system $A$ back to Alice through 
the second quantum channel. A direct transmission through the quantum channel 
can be replaced by quantum teleportation which uses a previously shared entanglement to 
transmit an unknown state of the system in an actual time only by local operations and classical communication (LOCC) \cite{Bennett93}. 
Can be this straightforward strategy based on a pair of quantum channels simplified ?
It was discovered that to implement basic CNOT gate between two qubits at a distance it is sufficient to utilize
only a single quantum channel \cite{CNOT}. A smaller number 
of the shared quantum channels substantially simplifies the 
implementation of the interaction at a distance and also increases a resistance to a 
decoherence effect owing to an imperfection in real quantum channels. 

In continuous-variable (CV) quantum information processing based only on Gaussian operations (for a review see Ref.~\cite{CV}), 
the QND coupling between two linear harmonic oscillators (LHOs) can provide a basic interaction for building any CV interaction 
between the multiple LHOs. 
Utilizing the CV QND coupling plus local unitary operations (such as squeezers, phase shifters and displacements) and the measurements 
of position and momentum, an arbitrary Gaussian operation with the continuous variables can be implemented \cite{Lloyd99}. 
Whereas in the experiments with discrete variables a strong nonlinear coupling between two qubits 
is required to deterministically implement the basic CNOT gate, 
the CV QND coupling is based on much more feasible Gaussian interaction between the LHOs. 
In quantum optics, the QND interaction between light pulses was widely investigated in many previous experiments \cite{QNDexp}. 
Recently, the QND interaction has been experimentally observed between the polarization of light pulses and collective spin of 
huge atomic samples \cite{Polzik}. 

Thus to distributively perform a complex experiment with the continuous variables 
an efficient realization of the CV QND interaction at a distance should be found.
Analogically, a straightforward realization of a CV interaction at a distance requires two CV quantum channels.
In this paper, two feasible implementations of the CV QND interaction at a distance requiring only a {\em single} 
quantum channel are proposed and the influence of Gaussian noise in the quantum channel is analyzed. 
Based on the previous experimental achievements \cite{QNDexp,Polzik}, 
two feasible experiments which can be used to demonstrate the CV QND interaction at a distance are discussed. 
  
We assume that Alice and Bob locally dispose two LHOs $A$ and $B$ having the coordinate and momentum variables 
$X_{A},P_{A}$ and $X_{B},P_{B}$, respectively. They satisfy the standard commutation relations $[X_{j},P_{k}]=i\delta_{jk}$, $j,k=A,B$. 
An example of the CV QND interaction at a distance between two LHOs $A$ and $B$ 
can be described by the following interaction 
Hamiltonian $H_{I}=\hbar\kappa X_{A}P_{B}$ \cite{QNDteo}, for which the coordinate and momentum operators transform 
in Heisenberg picture as 
\begin{eqnarray}\label{QND1}
X'_{A}&=&X_{A},\hspace{0.1cm}
P'_{A}=P_{A}-gP_{B},\nonumber\\
X'_{B}&=&X_{B}+gX_{A},\hspace{0.1cm}
P'_{B}=P_{B}
\end{eqnarray}
where $g=\kappa t$ is the gain of the QND interaction at a distance and $t$ is the interaction time. 
Physically, information encoded in a non-demolishing variable -- 
the $X_{A}$ coordinate, is transmitted to the $X_{B}$ coordinate however at the expense of 
additional noise in the momentum $P_{A}$. Note that 
analogical results, as will be discussed below,  can be also obtained for the other QND couplings.

If Alice and Bob wish to accomplish this QND interaction at a distance they can straightforwardly use 
two CV quantum channels. First, we demonstrate that if both Alice and Bob locally possess two identical QND interactions 
(\ref{QND1}) and in addition, Bob has an appropriate single-mode squeezed state 
they need to share only a {\em single} quantum channel to achieve (\ref{QND1}) at a distance with an arbitrary accuracy. 
The scheme is depicted in Fig.~1. We assume that Bob has an auxiliary 
LHO $C$ described by the quadrature variables $X_{C},P_{C}$. First, Bob locally 
applies the following sequence of operations: the phase shift about $\pi$ on LHO $C$ 
which transforms $X_{C}\rightarrow -X_{C}$ and $P_{C}\rightarrow -P_{C}$,  
the QND interaction (\ref{QND1}) with the gain $G$ between LHOs $B,C$ and once 
more the phase shift about $\pi$ on the LHO $C$. After this procedure Bob  
obtains in fact the following QND interaction between LHOs $B,C$ 
\begin{eqnarray}\label{QND2} 
X'_{C}&=&X_{C},\hspace{0.1cm}
P'_{C}=P_{C}+GP_{B},\nonumber\\
X'_{B}&=&X_{B}-GX_{C},\hspace{0.1cm}
P'_{B}=P_{B}
\end{eqnarray}
from the interaction (\ref{QND1}). 
Subsequently, Bob transmits the LHO $C$ through a perfect quantum channel 
to Alice and she locally couples it with LHO $A$ by the QND interaction (\ref{QND1}) 
with the gain $G$. After this second QND interaction, Alice and Bob in fact jointly perform the following transformation 
on LHOs $A,B,C$:
\begin{eqnarray}\label{inter} 
X'_{A}&=&X_{A},\hspace{0.1cm}
P'_{A}=P_{A}-G^{2}P_{B}-GP_{C},\nonumber\\
X'_{B}&=&X_{B}-GX_{C},\hspace{0.1cm}
P'_{B}=P_{B},\nonumber\\
X'_{C}&=&X_{C}+GX_{A},\hspace{0.1cm}
P'_{C}=P_{C}+GP_{B}.
\end{eqnarray}
Then Alice measures the coordinate of LHO $C$ with the result of $\bar{x}$ and 
this result is classically communicated to Bob. According to it, he performs 
a coordinate displacement $D_{X}$ on his LHO $B$, specifically $X'_{B}=X_{B}+G\bar{x}$. 
Consequently, the total transformation of the variables of LHOs $A,B$ in Heisenberg picture reads
\begin{eqnarray}\label{final} 
X'_{A}&=&X_{A},\hspace{0.1cm}
P'_{A}=P_{A}-G^{2}P_{B}-GP_{C},\nonumber\\
X'_{B}&=&X_{B}+G^{2}X_{A},\hspace{0.1cm}
P'_{B}=P_{B},
\end{eqnarray}
approaches the ideal QND interaction (\ref{QND1}) at a distance with the gain $g=G^2$ as the initial fluctuations 
of zero-mean variable $P_{C}$ tends to zero. It can be accomplished if the LHO $C$ is prepared in a 
sufficiently squeezed vacuum state in the momentum variable. Thus the distant users can build the QND interaction (\ref{QND1})
between them with an arbitrary accuracy employing only a {\em single} quantum channel and classical one-way communication \cite{note1}. 
Note, that the proposed scheme can be optimized if Alice and Bob have at their disposal 
the local QND couplings with arbitrary generally different gains $G_{A}$ and $G_{B}$. 
The gain $g$ of the QND coupling at a distance is $g=G_{A}G_{B}$ 
and the additive noisy term in Eqs.~(\ref{final}) is given by operator $-G_{A}P_{C}$. Then, to implement 
the QND coupling with the gain $g$ at a distance, it is better to use a smaller $G_{A}$ with a larger $G_{B}=g/G_{A}$.   

%%%%%%%%%%%%%%%%%%%%%%%%%%%%%%%%%%%%%%%%%%%%%%%%%%%%%%%%%%%%%%%%%%%%%%%%%%%%%%%%%%%%%%%%%%%
\begin{figure}
\vspace{1cm}
\centerline{\psfig{width=7.0cm,angle=0,file=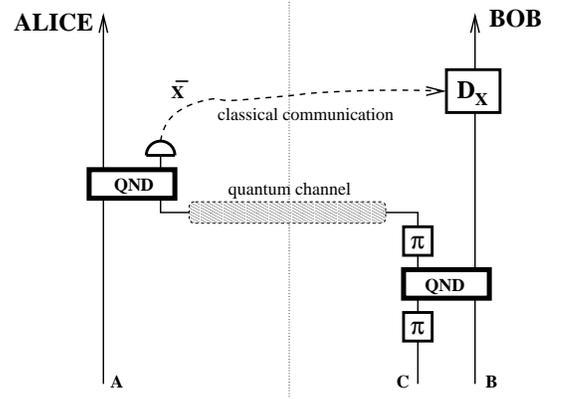}}
\caption{The scheme of CV QND interaction at a distance using a 
single perfect quantum channel and single-way classical communication.}
\end{figure}
%%%%%%%%%%%%%%%%%%%%%%%%%%%%%%%%%%%%%%%%%%%%%%%%%%%%%%%%%%%%%%%%%%%%%%%%%%%%%%%%%%%%%%%%%%%%
%%%%%%%%%%%%%%%%%%%%%%%%%%%%%%%%%%%%%%%%%%%%%%%%%%%%%%%%%%%%%%%%%%%%%%%%%%%%%%%%%%%%%%%%%%%
\begin{figure}
\vspace{1cm}
\centerline{\psfig{width=7.0cm,angle=0,file=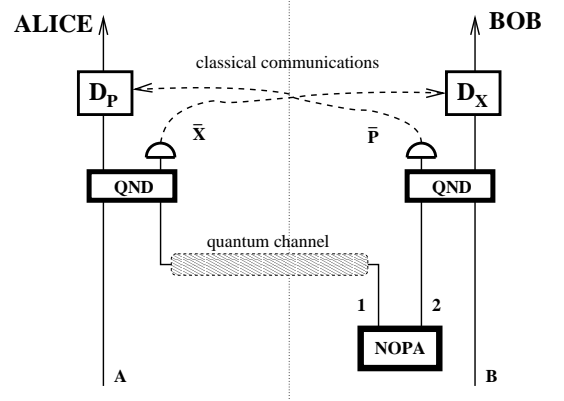}}
\caption{The scheme of CV QND interaction at a distance using apriori shared entanglement and 
two-way classical communication.}
\end{figure}
%%%%%%%%%%%%%%%%%%%%%%%%%%%%%%%%%%%%%%%%%%%%%%%%%%%%%%%%%%%%%%%%%%%%%%%%%%%%%%%%%%%%%%%%%%%%

However, the proposed setup on Fig.~1 still utilizes the quantum communication between 
Alice and Bob in an actual time of QND interaction at a distance.  
Now we demonstrate that using an apriori shared CV entangled state through the quantum channel 
the quantum communication in the actual time can be eliminated.
We assume in addition to LHOs $A,B$ a pair of LHOs $1,2$ described by the coordinate and momentum operators 
$X_{1},X_{2}$, $P_{1},P_{2}$. The LHOs $1,2$ are prepared in an entangled state exhibiting reduced fluctuations in the 
commuting variables $X_{1}+X_{2}$ and $P_{1}-P_{2}$. 
Mixing of two orthogonally squeezed vacuum state on balanced beam splitter or spontaneous emission in 
non-degenerate optical parametric amplifier (NOPA) are appropriate sources of such a state. 
The source is for example located on Bob's side as shown in Fig.~2. 
The QND interaction at a distance with the gain $g$ can be performed as follows. First, both Alice and Bob 
locally perform the QND interactions (\ref{QND1}) with the gain $G$ between the pairs of LHOs $A,1$ and $B,2$
and subsequently, Alice measures the coordinate variable $X_{1}$ on the LHO $1$ and Bob 
detects the momentum variable $P_{2}$ on the LHO $2$. After these measurements, 
Alice and Bob use two-way classical communication to mutually exchange the 
measured values. Then Bob performs the coordinate displacement $X'_{B}=X_{B}+G\bar{x}_{1}$ according to Alice's result
whereas Alice applies the momentum displacement $P'_{A}=P_{A}+G\bar{p}_{2}$ according to Bob's result. 
After the displacements, the following transformation in Heisenberg picture reads resulting
\begin{eqnarray}\label{final2}
X'_{A}=X_{A},\hspace{0.2cm}P'_{A}=P_{A}-G^{2}P_{B}-G(P_{1}-P_{2}),\nonumber\\
X'_{B}=X_{B}+G^{2}X_{A}+G(X_{1}+X_{2}),\hspace{0.2cm}P'_{B}=P_{B},  
\end{eqnarray}
and it approaches the QND interaction (\ref{QND1}) at a distance exhibiting the gain $g=G^{2}$ with arbitrary
accuracy if the entangled state of LHOs $1,2$ has substantially reduced fluctuations in the commuting 
variables $X_{1}+X_{2}$ and $P_{1}-P_{2}$. 
Thus the resource of entanglement suitable also for the CV teleportation experiment 
\cite{CVtele} can be used also to implement the QND interaction at a distance. 
Note, a swapping two QND interactions at a distance between laboratories ${\cal A},{\cal C}$ can be performed if two entanglement 
states shared between the laboratories ${\cal A},{\cal B}$ and ${\cal B},{\cal C}$ are swapped in the laboratory ${\cal B}$, 
utilizing the entanglement swapping scheme in Ref.~\cite{swapping}. 

In a summary, the QND interaction at a distance can be implemented with an 
arbitrary accuracy, using only the single sufficiently entangled state shared between Alice and Bob 
and classical two-way communication.
%classical versus quantum
We can compare this result with a classical realization of the QND operation at a distance which does not stem 
from the shared entanglement. The LHOs $1,2$ in Eq.~(\ref{final2}) are assumed to be prepared in the ground state 
for the classical analog. Thus, variance of the noise added in both the transformed variables is 
$2G^{2}V_{0}$, where $V_{0}$ is the variance of coordinate and momentum in the ground state. 
This noise excess deteriorates quantum features of the QND interaction at a distance and can be only reduced by
an entanglement shared between Alice and Bob. Any classical realization of the QND interaction at distance is 
not able to generate the entanglement between Alice and Bob from separable states at the input and thus, 
a small amount of the entanglement shared between Alice and Bob in the proposed method qualitatively exceeds the classical 
realization.
%comparison 1and 2
Further let us compare both the proposed methods depicted in Figs.~1,2. In the first one in Fig.~1, information 
about the state of the LHO $B$ is partially transmitted to Alice 
through the quantum channel contrary to the second one in Fig.~2, where all the needed information about LHOs $A$ 
and $B$ is mutually exchanged only by classical two-way communication. 
From the point of view of quantum noise associated with a finite CV squeezing or 
finite CV entanglement, the additional noise represented by the term $GP_{C}$ in Eqs.~(\ref{final}) is asymmetrically 
induced only in a single variable $P'_{A}$ in the first method
whereas the noises arising from the terms $G(X_{1}+X_{2})$ and $G(P_{1}-P_{2})$ 
symmetrically affect both variables $P'_{A}$ and $X'_{B}$ in the second method.   

%nonlocal gates with imperfect channel
A damping in the quantum channel enlarges a noise in the variables $X_{1}+X_{2},P_{1}-P_{2}$ in practice, 
it can lead to a decrease of quality of the realization of the QND interaction at a distance.  
As a typical example, the lossy channel with a known transmitivity $T\not= 0$ coupled to the vacuum environmental modes can be assumed. 
This lossy channel can be at least partially improved for our purpose if 
a phase-insensitive amplifier with a gain compensating the loss of energy in the channel is inserted in front of
the channel. Resulting noisy channel can be described by the following transformation: 
\begin{equation}\label{channel}
X'=X+\sqrt{1-T^2}{\cal X},\hspace{0.3cm} P'=P+\sqrt{1-T^2}{\cal P}
\end{equation} 
of the coordinate and momentum operators, 
where ${\cal X},{\cal P}$ are the noise operators having zero mean values. They describe a noise produced by 
the vacuum environmental modes and pre-amplification process. 
Thus, any Gaussian state propagating through this noisy channel has unchanged mean 
values of both $X,P$ quadratures and only the variances in both quadratures are equally enhanced. 
From this follows that a fidelity of the transmission is state-independent.  

Assuming this noisy channel (\ref{channel}) between Alice and Bob,  
Eqs.~(\ref{final}) are substituted for the method depicted in Fig.~1 by 
\begin{eqnarray}\label{noise1}
X'_{A}&=&X_{A},\hspace{0.3cm}P'_{B}=P_{B}\nonumber\\
P'_{A}&=&P_{A}-G^{2}P_{B}-GP_{C}-G\sqrt{1-T^2}{\cal P},\nonumber\\
X'_{B}&=&X_{B}+G^{2}X_{A}+\sqrt{1-T^2}{\cal X}, 
\end{eqnarray}
whereas for the method depicted in Fig.~2, 
the relations (\ref{final2}) are replaced by the following ones
\begin{eqnarray}\label{noise2}
X'_{A}&=&X_{A},\hspace{0.3cm} P'_{B}=P_{B},\nonumber\\
P'_{A}&=&P_{A}-G^{2}P_{B}-G(P_{1}-P_{2})-G\sqrt{1-T^2}{\cal P},\nonumber\\
X'_{B}&=&X_{B}+G^{2}X_{A}+G(X_{1}+X_{2})+G\sqrt{1-T^2}{\cal X}.\nonumber\\ 
\end{eqnarray}
To compare both the methods from the point of view of the noise generated only by the channel, 
the transmitivity $T$ is fixed and it is considered that the single-mode squeezing or 
two-mode squeezing can be generate at will. 
Thus, a sufficient squeezing ensures that noise operator $GP_{C}$ in Eqs.~
(\ref{noise1}) and the operators $G(P_{1}-P_{2})$ and $G(X_{1}+X_{2})$  
in Eqs.~(\ref{noise2}) can be neglected and only the noise 
effects arising from the noisy channel are compared. 
Then, the method in Fig.2 is less noisy for QND gain $G<1$ whereas 
the method in Fig.~1 is better for $G>1$. Both the methods are equivalent for $G=1$. 
Note, assuming independent QND gains $G_{A},G_{B}$, that the method depicted in Fig.~1 can be optimized also if 
the quantum channel is lossy, analogically as has been discussed above. 

We can also compare both the methods with the straightforward two-way teleportation strategy 
through the noisy channel (\ref{channel}). In the two-way teleportation strategy,
Bob teleports his state to Alice through the first noisy channel, Alice locally performs 
the QND interaction with gain $G^2$ and then teleports one of the output states back to Bob through the second noisy channel. 
Using this strategy, the transformations for the coordinate and momentum operators are 
\begin{eqnarray}\label{noise3}
X'_{A}&=&X_{A},\hspace{0.3cm}P'_{B}=P_{B}+{\cal L}_{P}^{(1)}+{\cal L}_{P}^{(2)}\nonumber\\
P'_{A}&=&P_{A}-G^{2}P_{B}-G^{2}{\cal L}_{P}^{(1)},\nonumber\\
X'_{B}&=&X_{B}+G^{2}X_{A}+{\cal L}_{X}^{(1)}+{\cal L}_{X}^{(2)}, 
\end{eqnarray}
where ${\cal L}_{X}^{(i)}=\sqrt{1-T^2}{\cal X}^{(i)}+(X_{1}^{(i)}+X_{2}^{(i)})$ and 
${\cal L}^{(i)}_{P}=\sqrt{1-T^2}{\cal P}^{(i)}+(P_{1}^{(i)}-P_{2}^{(i)})$, $i=1,2$ 
are the noise operators jointly arising from both the noisy channels and imperfect teleportations. 
A noise induced by the operators $(X_{1}^{(i)}+X_{2}^{(i)})$ and $(P_{1}^{(i)}-P_{2}^{(i)})$ 
is related to a finite entanglement in the quantum teleportations, whereas the 
channel noise is represented by the operators ${\cal X}^{(i)}$ and ${\cal P}^{(i)}$. 
As an example, the implementation of the QND interaction at a distance 
with unit gain $g=1$ can be simply discussed. If a sufficiently  
large entanglement shared between Alice and Bob in the two-way teleportation scheme is assumed then an 
influence of the noise operators $X_{1}^{(i)}+X_{2}^{(i)}$, 
$P_{1}^{(i)}-P_{2}^{(i)}$ can be neglected. 
Then, the noise operators arising from the channel in 
Eqs.~(\ref{noise1},\ref{noise2}) 
have a smaller influence in the proposed implementation than the equivalent ones in Eqs.~
(\ref{noise3}) for the two-way teleportation scheme.   

%Application in optical and atom-optical experiments 
In quantum optics, the QND interactions (\ref{QND2}) were previously experimentally implemented 
between two modes of optical fields with orthogonal polarizations.   
The setup consists of a non-degenerate parametric amplifier with a suitable gain sandwiched by two $\lambda/2$ wave 
plates with optical axis at specific angle with respect 
to orthogonal polarization \cite{QNDexp}. Here the coordinate and momentum variables 
$X$ and $P$ are represented by two complementary quadratures of a particular mode of the optical field. 
Using the proposed methods, an all-optical QND interaction at a distance can be realized, 
according to scheme in Fig.~1 using single-mode squeezed state produced by the parametric amplification 
or according to scheme in Fig.~2, by sharing a CV entangled state generated 
by mixing two orthogonally squeezed single-mode states on a balanced beam splitter \cite{NOPA}. 
Recently, a novel hybrid version of the QND coupling between the polarization of bright 
light pulses and collective spin of a sample of $10^{12}$ cesium atoms
has been also demonstrated \cite{Polzik}. In these experiments, 
the coordinate and momentum operators are effectively two projections $J_{y}$ and $J_{z}$ 
of the collective spin of the atomic sample and Stokes operators $S_{y}$ and $S_{z}$
of the linearly polarized bright beam.  
A version of the QND interaction occurring between the atomic sample and the light beam
can be effectively described as the transformation $S'_{y}=S_{y}+G J_{z}$, $J'_{y}=J_{y}+G S_{z}$ whereas 
the components $S_{z}$ and $J_{z}$ are unchanged. Using entangled   
polarization state of light \cite{NOPApol}, the 
QND measurement with the gain $G^{2}$ between two distant atomic samples can be demonstrated. 
 
In this paper, using only a single quantum channel and classical communication, 
two feasible experimental implementations of the CV QND interaction at a distance are proposed. The 
influence of the Gaussian noisy quantum channel in both the implementations is discussed. They can be directly implemented in 
the recent quantum optical experiments in which the CV QND interactions were previously 
demonstrated \cite{QNDexp,Polzik}. An implementation of the interactions at a distance can open a way toward a joint
experiment between the distant laboratories possessing different experimental techniques.  
For example, if Alice's laboratory works on the experiment with the light pulses and Bob's laboratory with the atomic samples 
they can perform a joint experiment without a necessity to move an experimental apparatus from one laboratory to the other one.

%%%%%%%%%%%%%%%%%%%%%%%%%%%%%%%%%%%%%%%%%%%%%%%%%%%%%%%%%%%%%%%%%%%%%%%%%%%%%%

\medskip
\noindent {\bf Acknowledgments}
I acknowledge fruitful discussions with 
Jarom\'ir Fiur\' a\v sek, Ladislav Mi\v sta, Petr Marek. 
The work was supported by the grant 202/03/D239 of the Czech Grant Agency and 
the project LN00A015 of the Ministry of Education of Czech Republic and by the EU grant under QIPC project
IST-1999-13071 (QUICOV). 

%%%%%%%%%%%%%%%%%%%%%%%%%%%%%%%%%%%%%%%%%%%%%%%%%%%%%%%%%%%%%%%%%%%


\begin{thebibliography}{99}
\bibitem{note1}
Note that Alice and Bob also 
need to share a local oscillator to perform 
an efficient homodynne detection of the quadrature variable.
\end{thebibliography}
\end{document}